\documentclass[journal]{IEEEtran}
\usepackage{amssymb}
\usepackage{graphicx}
\usepackage{graphics}
\usepackage{array} 
\usepackage{ctable}
\usepackage{multirow}
\usepackage[footnotesize]{subfigure}
\usepackage{upgreek}
\usepackage{cite}

\ifCLASSINFOpdf
\else
\fi
\begin{document}
\title{A Time-Frequency Perspective on \\ Audio Watermarking}
\author{Haijian~Zhang
\thanks{H. Zhang is with Signal Processing Laboratory, School of Electronic Information, Wuhan University, China.}
}

\markboth{Time-frequency analysis and its applications}
{}
\maketitle

\begin{abstract}
Existing audio watermarking methods usually treat the host audio signals of a function of time or frequency individually, while considering them in the joint time-frequency (TF) domain has received less attention. This paper proposes an audio watermarking framework from the perspective of TF analysis. The proposed framework treats the host audio signal in the 2-dimensional (2D) TF plane, and selects a series of patches within the 2D TF image. These patches correspond to the TF clusters with minimum averaged energy, and are used to form the feature vectors for watermark embedding. Classical spread spectrum embedding schemes are incorporated in the framework. The feature patches that carry the watermarks only occupy a few TF regions of the host audio signal, thus leading to improved imperceptibility property. In addition, since the feature patches contain a neighborhood area of TF representation of audio samples, the correlations among the samples within a single patch could be exploited for improved robustness against a series of processing attacks. Extensive experiments are carried out to illustrate the effectiveness of the proposed system, as compared to its counterpart systems. The aim of this work is to shed some light on the notion of audio watermarking in TF feature domain, which may potentially lead us to more robust watermarking solutions against malicious attacks.  
\end{abstract}

\begin{IEEEkeywords}
 Audio watermarking, Feature domain watermarking, Time-frequency domain watermarking, Short-time Fourier transform, Imperceptibility, Robustness.
\end{IEEEkeywords}

\section{Introduction}

With the rapid development of modern communication and multimedia technologies, the dissemination and processing of digital multimedia products are becoming more and more popular, which inevitably gives rise to a variety of piracy and infringement issues. Watermarking techniques have received significant research attention as a means to efficiently protect the copyright of digital multimedia product \cite{Cox2007}. While watermarks can be embedded into media formats including but not limited to document, image, audio, and video, in this paper, we focus on watermarking for audio signals, which are functions of time.

Recently, the authors in \cite{Hua2016_SP} reviewed the research, development, and commercialization achievements of digital audio watermarking technology for the past twenty years. Generally, the existing audio watermarking techniques could be classified according to the domains in which the watermarks are embedded. More specifically, time domain methods either modify the raw audio samples frame by frame \cite{Nishimura2012,Hua2015_TASLP,Hua2015_TIFS}, or change the histogram of host audio signals \cite{Xiang2007_TM}. On the other hand, transform domain methods, which have received much more research attention, can be classified into spread spectrum (SS) \cite{Cox1997_TIP,Kirovski2003_TSP,Malvar2003_TSP,Xiang2015_TASLP}, patchwork \cite{Xiang2014_TASLP}, quantization index modulation (QIM) \cite{Lei2013_TASLP}, and a special case based on over-complete transform dictionaries \cite{Cancelli2009,Hua2016_SPL}.

It could be seen from the literature that although audio watermarking solutions have been extensively studied in time or a transform domain individually, less efforts have been devoted to the case in which the host audio signal is analyzed and represented jointly in time-frequency (TF) domain, based on the well established TF analysis techniques (transforms) \cite{SejdiDjurovi2009,StankoviKrishnan2010,Boashash2016,Flandrin2018}.  TF analysis is a generalization of Fourier analysis for the case when the signal frequency characteristics are time-varying. Since many practical signals of interest, such as speech and music, have varying frequency characteristics, TF analysis has a broad scope of applications, and one of the most basic forms of TF analysis is the short-time Fourier transform (STFT) \cite{haijianDSP2016,haijianSC2017}, while more sophisticated techniques have also been developed \cite{Zhang2015_TAES,Zhang2016_SPL,Zhang2020_TIFS}. In \cite{Stankovic2008,Orovic2010}, the authors proposed two efficient approaches to speech watermarking based on the STFT and the S-method \cite{Stankovic1994}. 

\begin{figure*}[t!]
	\centering
	\subfigure[]{\includegraphics [width=4.5in]{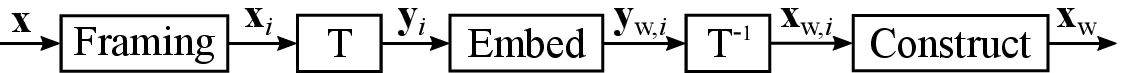}}
	\subfigure[]{\includegraphics [width=4.5in]{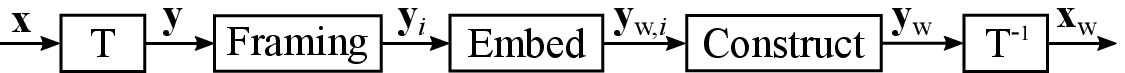}}
	\subfigure[]{\includegraphics [width=5.2in]{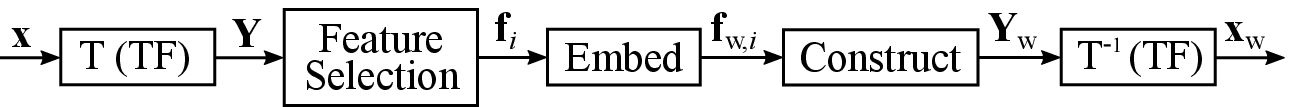}}
	\caption{Commonly used and the proposed watermark embedding schemes: (a) Transform after framing. (b) Transform before framing. (c) Proposed scheme.}\label{Schemes}
\end{figure*}

In this paper, we discuss audio watermarking from the perspective of time-frequency analysis, and propose an audio watermarking framework based on the fact that audio signals are a function of time. Specifically, we propose to embed the watermark signal into a set of low-energy points in the TF representation, which correspond to noise-only or silent segments in audio signals. The selected points can form non-overlapping 2-dimensional (2D) feature frames, each of which is composed of TF domain samples across multiple time and frequency bins. To achieve this purpose, a method to automatically determine these low-energy TF positions is introduced, and the energy invariance before and after watermark embedding is exploited. The proposed scheme only modifies a few frames within the feature space, while other frames are kept intact. The imperceptibility property of the watermarking system could hence be improved in that the host audio signal containing strong audio content is less modified. Therefore, while the robustness against host signal interference could be ensured via the use of improved spread spectrum (ISS) method \cite{Malvar2003_TSP}, the inability of ISS method to control imperceptibility is remedied by the proposed localized watermark embedding. Furthermore, the proposed system enjoys improved robustness against a series of signal processing attacks including adding noise, amplitude scaling, and lossy compressions, thanks to the appropriately designed 2D embedding frames. In general, the proposed framework could be considered as a design for feature domain audio watermarking, in which the features correspond to the appropriately selected embedding locations. 

We concretize the above framework via the realizations of the basic STFT and a similarly formed short-time cosine transform (STCT), with both SS and ISS embedding and extraction mechanisms. Conventional SS and ISS schemes with a uniform embedding rule across different frequency bands are also implemented for comparison. Extensive experiments are carried out to evaluate the proposed framework and demonstrate its performance advantages.


\section{Audio Watermarking in TF Feature Domain}
\subsection{General Frameworks}
We use the following notations in this paper. The host audio signal is denoted by vector $\mathbf{x}\in\mathbb{R}^{N\times 1}$, and its $i$th frame after time domain non-overlapping framing is $\mathbf{x}_i\in\mathbb{R}^{M_0 \times 1}$, where $N$ is the number of samples of the host signal, $M_0$ is the number of samples per frame, and $i\in\{0,1,\ldots,\lceil {N/M_0} \rceil-1\}$, where $\lceil {\cdot} \rceil$ is the ceiling function. Similarly, the host audio signal in transform domain is denoted by $\mathbf{y}$ with the same length as $\mathbf{x}$, and the $i$th frame of $\mathbf{y}$ after transform domain non-overlapping framing is $\mathbf{y}_i$. We use subscript $\{\cdot\}_{\textnormal{w}}$ to denote watermarked version of a signal, thus the representations of watermarked signal in time and transform domain, and in terms of frame and ensemble, are denoted by $\mathbf{x}_{\textnormal{w},i}$, $\mathbf{x}_{\textnormal{w}}$, $\mathbf{y}_{\textnormal{w},i}$, and $\mathbf{y}_{\textnormal{w}}$, respectively. Since this paper mainly utilizes SS based watermark embedding and extraction mechanisms, the corresponding spreading sequence is a pseudo-random noise sequence $\mathbf{p}\in\{+1,-1\}^{L\times 1}$. In conventional full spectrum SS settings, we have $L=M_0$ so that the spreading sequence could be additively embedded into host signal frames.

Audio watermark embedding in transform domain could be carried out under two basic schemes, i.e., transform after and before framing, which are shown in Fig. \ref{Schemes} (a) and (b). Transform after framing is the most widely adopted processing flow in the existing literature, which is summarized in \cite{Hua2016_SP}. On the other hand, host audio signal has also been considered as a single frame to calculate its corresponding transform domain representations, directly obtaining $\mathbf{y}$ from $\mathbf{x}$. Good examples of such works could be seen in \cite{Xiang2015_TASLP,Xiang2014_TASLP,Kang2016_TM}. In this paper, we propose an alternative watermark embedding framework as depicted in Fig. \ref{Schemes} (c), which is similar to the transform before framing case. Specifically, instead of calculating the transform $\mathbf{y}$, we fist obtain the TF representation of the host signal $\mathbf{x}$, denoted by matrix $\mathbf{Y}\in\mathbb{C}^{M_0 \times \lceil {N/M_0} \rceil }$ composed of both time and frequency bins. In this way, the proposed framework differs from most of existing ones by considering watermark embedding based on a 2D TF image. Furthermore, the TF domain feature, denoted by $\mathbf{f}_i\in\mathbb{C}^{W^2 \times 1}$ (where $W$ is a window dimension), is selected as the patches with low-energy values, which correspond to noise-only or silent locations. One of the advantages of using the proposed framework over the similar framework in Fig. \ref{Schemes} (b) is that the modification of host signal in one area will not affect the host signal in other areas, while for a system in Fig. \ref{Schemes} (b), any modification in transform domain samples will cause changes of all samples in time domain. In addition, since each of the selected feature vectors contains multiple time and frequency bins, the correlation of the host signal at both different time intervals and frequency ranges is considered for watermark embedding, which could lead to improved robustness against a series of processing and attacks. Details of the proposed framework are provided in next subsection. 

\begin{figure}[!t]
	\centering
	\includegraphics[width=2.5in]{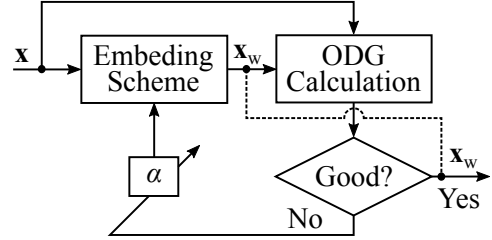}
	\caption{Heuristic tuning mechanism to control imperceptibility based on ODG. The dashed line is validated only if the decision is ``Yes'' (stopping criterion).}
	\label{Tune}
\end{figure}

\subsection{Watermark Embedding and Extraction Schemes}
In this subsection, STFT is used as an example for TF analysis. To ensure well controlled imperceptibility, heuristic tuning is incorporated in the embedding scheme, in which the objective difference grade (ODG) \cite{ODG} is utilized to quantify current imperceptibility condition, according to which the watermark embedding strength parameter $\alpha$ is adjusted according to the feedback of the ODG value. The ODG value is a real non-positive number in the intervals of $\{-4,-3,-2,-1,0\}$, with $0$ corresponding to imperceptible and -$4$ corresponding to very annoying. The heuristic tuning mechanism is depicted in Fig. \ref{Tune}. The proposed watermark embedding scheme is detailed as follows. 

\begin{enumerate}
\item Partition $\mathbf{x}$ into non-overlapping frames $\mathbf{x}_i$ with $M_0$ samples\footnote{Here, the frames could not be overlapped because otherwise the watermarked patches will affect multiple overlapped frames and the inverse transform would become unstable. This is slightly different from TF analysis literature. However, such a treatment will not cause performance degradation since we are not interested in the resolution or accuracy of TF analysis in the context of watermarking.}. Perform Hilbert transform on $\mathbf{x}_i$ to remove the symmetry of the frequency spectrum within $2\pi$ radius \cite{Grafakos2004_book}. For simplicity of notation, we still use $\mathbf{x}_i$ to denote the Hilbert transform output,
\begin{equation}\label{Hilbert}
{{\mathbf{x}}_i}\in\mathbb{C}^{M_0\times 1} \leftarrow \textnormal{Hilbert} \left( {{{\mathbf{x}}_i}} \right), i={0,1,\cdots,\lceil {N/M_0} \rceil-1},
\end{equation} 	
but note here $\mathbf{x}_i$ becomes complex quantities.

\item Compute the non-symmetric STFT of the host audio signal $\mathbf{x}$, and obtain the TF representation $\mathbf{Y}$. Specifically, perform fast Fourier transform (FFT) for each frame,
\begin{equation}
{{\mathbf{y}}_i} = {\mathbf{H}}{{\mathbf{x}}_i},
\end{equation}
where $\mathbf{H}$ is the orthonormal FFT matrix. Then we have
\begin{equation}
{\mathbf{Y}} = [{{\mathbf{y}}_0},{{\mathbf{y}}_1}, \ldots ,{{\mathbf{y}}_{\left\lceil {N/M_0} \right\rceil  - 1}}].
\end{equation}
Further, select low to middle frequency bins bounded by $f_1$ and $f_2$, e.g., $f1=60$ Hz, and $f_2=2800$ Hz, as the feasible watermark embedding region. The exact dimension, $M$, resulted from this process depends on $f_1$, $f_2$, the sampling frequency, and the length of FFT. We simply denote the refined 2D TF image as $\tilde{\mathbf{Y}}\in\mathbb{C}^{M\times \lceil {N/M_0} \rceil}$. Therefore, vertically, the $M$ samples correspond to a frequency region within $[f_1,f_2]$. 

\item Partition $\tilde{\mathbf{Y}}$ into square patches using a $W\times W$ window, and index the patches in raster scanning order. Usually, we have
\begin{equation}
W < M < M_0  \ll N.
\end{equation} 
For convenience, we further assume $W$ is chosen such that $W^2$ is divisible by both $M$ and ${\left\lceil {N/M_0} \right\rceil }$, hence there is no residual after partition, and $M\left( {\left\lceil {N/M_0} \right\rceil} \right)/{W^2}$ patches are obtained in total. 

\item Calculate the average energy of each patch. Denote each patch as $\mathbf{P}_j$, $j\in\{0,1,\ldots,M\left( {\left\lceil {N/M_0} \right\rceil  - 1} \right)/{W^2}-1\}$, then, the average energy is given by
\begin{equation}
{E_j} = \frac{1}{{{W^2}}}\sum\limits_{{m_1} = 0}^{W - 1} {\sum\limits_{{m_2} = 0}^{W - 1} {\left| {{{\mathbf{P}}_j}({m_1},{m_2})} \right|^2} }.
\end{equation}

\begin{figure}[t!]
	\centering
	\includegraphics[width=3.6in]{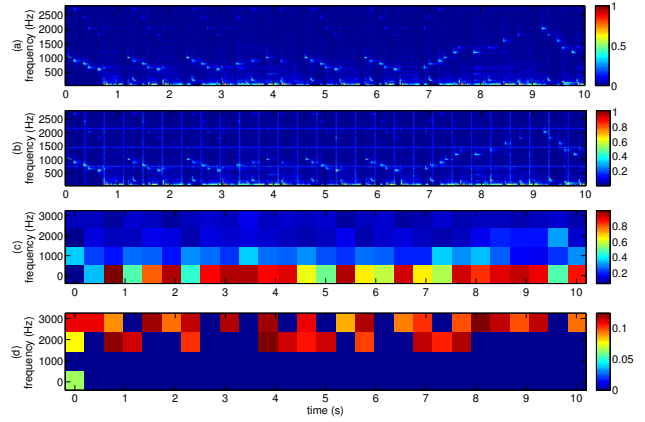}\\[-1em]
	\caption{Demonstration of the proposed STFT-based watermark embedding scheme. (a) STFT of host audio signal. (b) Partition of the 2D TF image. (c) Energy image of the patches of the 2D TF image. (d) Selected feature patches (red, payload size $P=32$).}
	\label{Embed}
\end{figure}

\item Sort $E_j$ in ascending order. According to the binary payload vector $\mathbf{w}\in\{+1,-1\}^{P\times 1}$, usually,
\begin{equation}
P<M\left( {\left\lceil {N/M_0} \right\rceil } \right)/{W^2},
\end{equation}
select the first $P$ patches with minimum average energies as features for watermark embedding. Vectorize the selected patches into feature vectors $\mathbf{f}_i\in\mathbb{C}^{W^2 \times 1}$, $i\in\{0,1,\ldots,P-1\}$. Next, the embedding order is from top to bottom and column-wise, i.e., the feature patches in the first column are first embedded from high frequency bands to low frequency bands, followed by patches in the second column, as so on.

\item For each feature vector ordered as above, generate the PN sequence $\mathbf{p}\in\{+1,-1\}^{W^2 \times 1}$ as the spreading code, and perform SS or ISS watermark embedding additively, i.e.,
\begin{equation}\label{EmbedFun}
{{\mathbf{f}}_{{\textnormal{w,}}i}} = {{\mathbf{f}}_i} + \left( {\alpha {\mathbf{w}}(i) - I\Upphi} \right){\mathbf{p}},
\end{equation}
where
\begin{equation}
\Upphi  \triangleq  \frac{{{\mathbf{f}}_i^T{\mathbf{p}}}}{{\left\| {\mathbf{p}} \right\|_2^2}},
\end{equation}
$\{\cdot\}^T$ is transpose operator, and $0<\alpha<1$ controls watermark embedding strength. For simplicity, parameter $I$ is a binary indicator, i.e., if $I=0$, then the scheme is based on SS, while if $I=1$, then the scheme is based on ISS. 

\item After embedding the payload $\mathbf{w}$, $\mathbf{Y}_\textnormal{w}$ is obtained by simply replacing its subset $\tilde{\mathbf{Y}}$ with $\tilde{\mathbf{Y}}_\textnormal{w}$. Then, perform inverse STFT according to the same framing rule as used in Step 1, and reorder the output to vector form and discard the imaginary part to obtain $\mathbf{x}_\textnormal{w}$. 

\item Calculate the ODG value according to $\mathbf{x}$ and $\mathbf{x}_\textnormal{w}$. Adjust parameter $\alpha$ according to a desired ODG level, i.e., if the ODG value is greater than the desired value (more imperceptible), then $\alpha$ could be slightly increased as long as the resultant ODG is within a tolerant distance from the desired value; if the ODG value is smaller than the desired value (less imperceptible), then $\alpha$ should be reduced accordingly. This process is shown in Fig. \ref{Tune}.

\end{enumerate}
The watermark embedding process is visualized in Fig. \ref{Embed}. At the receiving end, assuming an error-free channel, the extraction of the payload in terms of the detection of each embedded information bit is carried out as follows.
\begin{enumerate}

\item  Partition $\mathbf{x}_\textnormal{w}$ into non-overlapping frames $\mathbf{x}_{\textnormal{w},i}$ with $M_0$ samples, and perform Hilbert transform similar to (\ref{Hilbert}).

\item Compute the non-symmetric STFT of $\mathbf{x}_{\textnormal{w}}$, and obtain the TF representation $\mathbf{Y}_{\textnormal{w}}$. Specifically, perform FFT for each frame,
\begin{equation}
{{\mathbf{y}}_{\textnormal{w},i}} = {\mathbf{H}}{{\mathbf{x}}_{\textnormal{w},i}},
\end{equation}
then we have
\begin{equation}
{\mathbf{Y}_\textnormal{w}} = [{{\mathbf{y}}_{\textnormal{w},0}},{{\mathbf{y}}_{\textnormal{w},1}}, \ldots ,{{\mathbf{y}}_{\textnormal{w}, {\left\lceil {N/M_0} \right\rceil  - 1}}}].
\end{equation}
Further, according to $f_1$ and $f_2$, construct the sub-matrix $\tilde{\mathbf{Y}}_\textnormal{w}\in\mathbb{C}^{M\times \lceil {N/M_0} \rceil}$.

\item Partition $\tilde{\mathbf{Y}}_\textnormal{w}$ into square patches using a $W\times W$ window, and index the patches in raster scanning order.

\item Calculate the average energy of each patch. Denote each patch as $\mathbf{P}_{\textnormal{w},j}$, $j\in\{0,1,\ldots,M\left( {\left\lceil {N/M_0} \right\rceil  - 1} \right)/{W^2}-1\}$, then, the average energy is given by
\begin{equation}
{E_{\textnormal{w},j}} = \frac{1}{{{W^2}}}\sum\limits_{{m_1} = 0}^{W - 1} {\sum\limits_{{m_2} = 0}^{W - 1} {\left| {{{\mathbf{P}}_{\textnormal{w},j}}({m_1},{m_2})} \right|^2} } .
\end{equation}

\item Sort ${E_{\textnormal{w},j}}$ in ascending order, and find $P$ patches with least energy values. Vectorize these patches to form  $\mathbf{f}_{\textnormal{w},i}$ which are ordered column-wise and from top to bottom. The embedded information bit is estimated by the following function
\begin{eqnarray}\label{Extract}
\hat{\mathbf{w}}(i) & = & \textnormal{sgn} \left\langle {\Re({{\mathbf{f}}_{\textnormal{w},i}}),{\mathbf{p}}} \right\rangle /\left\| {\mathbf{p}} \right\|_2^2 \nonumber\\
 & = & \textnormal{sgn} \left( {\frac{{\Re({\mathbf{f}}_i)^T{\mathbf{p}} + \left( {\alpha {\mathbf{{w}}}(i) - I\Re({\Upphi}) } \right)\left\| {\mathbf{p}} \right\|_2^2}}{{\left\| {\mathbf{p}} \right\|_2^2}}} \right) \nonumber\\
 & = &  \textnormal{sgn} \left( {(1 - I)\Re({\Upphi})  + \alpha {\mathbf{{w}}}(i)} \right),
\end{eqnarray}
where $\Re\{\cdot\}$ denotes the real part.
\end{enumerate}
It can be seen from (\ref{Extract}) that the SS scheme ($I=0$) suffers from host signal interference $\Upphi$, while the ISS scheme ($I=1$) is able to remove the interference term in a closed-loop environment. Note that the above embedding and extraction schemes could be modified to other schemes by simply replacing the STFT with STCT or other transforms.

\subsection{Feature Invariance}
\begin{figure}[!t]
	\centering
	\subfigure[\small Top: STFT. Mid.: Feature patches. Bot.: Recovered patches.]{\includegraphics [width=3.7in]{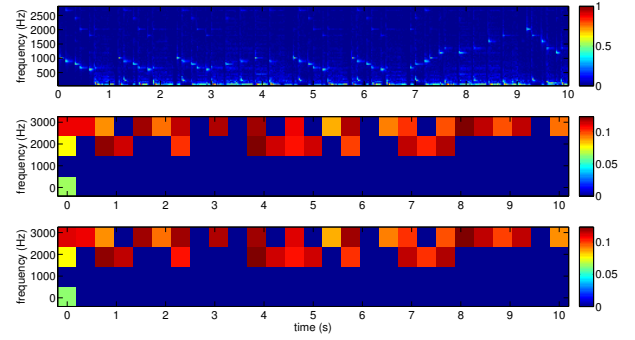}}\\[-.5em]
	\subfigure[\small Top: STCT. Mid.: Feature patches. Bot.: Recovered patches.]{\includegraphics [width=3.7in]{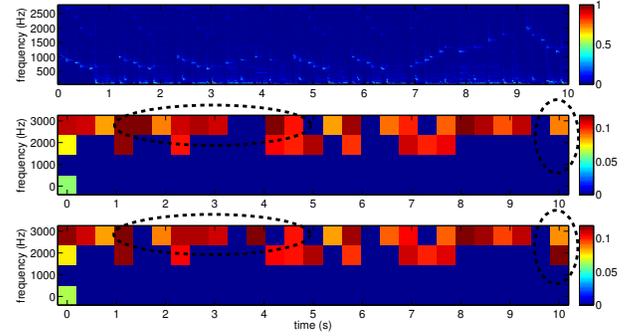}}\\[-.5em]
	\caption{Detection results of watermark positions via (a) STFT and (b) STCT.}
	\label{FeatureRecover}
\end{figure}

In this subsection, we address an important issue to validate the proposed system, i.e., feature invariance before and after watermark embedding. It can be seen from the embedding function (\ref{EmbedFun}) that the energy of watermarked feature vector will be altered. Therefore, after the whole embedding process, the energy distribution of the feature patches in $\mathbf{Y}_\textnormal{w}$, at least the $P$ patches with least energy levels, should still have least energy levels. To study the feature recovery property of the proposed system under different TF transforms, the recovery results using a sample audio clip in a closed-form environment are depicted in Fig. \ref{FeatureRecover}, where the audio clip has a duration of $10$ seconds, and $P=32$. It could be seen that STFT based method is more suitable for the proposed framework. The reason behind is that DCT tends to compact the signal's energy into smaller frequency band and also de-correlate the signal in frequency domain, making the small energy regions more ambiguous in terms of energy difference. Therefore, in the sequel, we will only consider STFT as the TF analysis tool in the following experiments. Extensive experimental results will be provided later to demonstrate how additive noise and other processing attacks affect the effectiveness of the proposed framework.

It is worth noting that there actually exists another solution for the requirement of feature invariance, which could be obtained using an indexing array pointing to $P$ patches randomly to identify which patches are watermarked. This array could be considered as a private key shared among the authority and trusted parties. \textbf{Note that the system proposed in the previous subsection is strictly a blind watermarking scheme whose watermark extraction does not require any auxillary information, but if the random indexing array is introduced, then this array, serving as a key, should be transmitted to authorized receivers via some secure channels.} The study of random index key based TF feature domain watermarking is noted here for future research attention.

\begin{figure}[!t]
	\centering
	\subfigure[DR versus DWR.]{\includegraphics [width=2.8in]{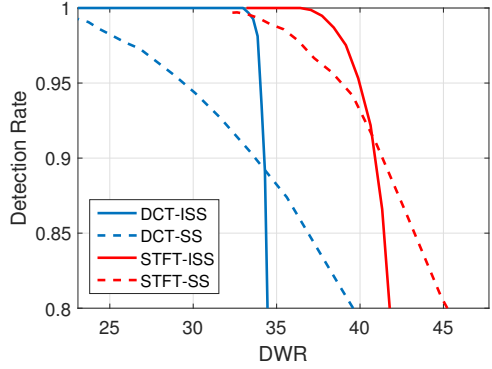}}\\[-.5em]
	\subfigure[DR versus ODG.]{\includegraphics [width=2.8in]{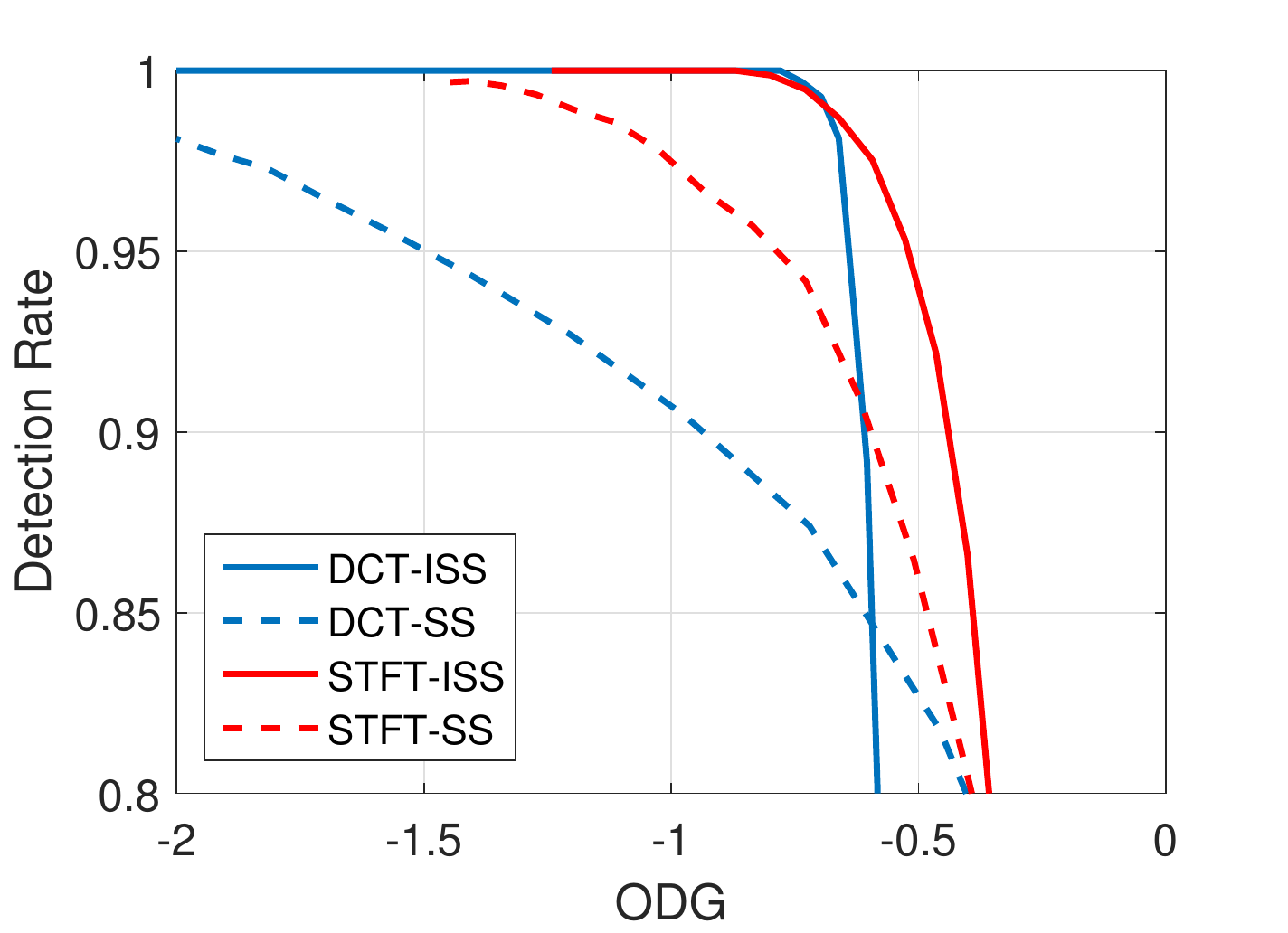}}\\[-.5em]
	\caption{Watermark detection rates averaged by running $3$ audio samples: (a) Detection Rate versus DWR. (b) Detection Rate versus ODG.}
	\label{Imp}
\end{figure}

\section{Evaluations and Experimental Results}
In this section, we carry out extensive experiments to evaluate the proposed framework in terms of imperceptibility and robustness. The imperceptibility is measured quantitatively by document-to-watermark ratio (DWR) and ODG respectively.
For comparison, the counterpart system based on DCT and the scheme in Fig. \ref{Schemes} (b) is also implemented. Therefore, the experiments and comparisons will be conducted on four systems, i.e., STFT-SS, STFT-ISS, DCT-SS, and DCT-ISS, respectively. Some measurement metrics are defined as follows. First, the DWR is given by
\begin{equation}
\textnormal{DWR} = 10{\log _{10}}\frac{{\left\| {\mathbf{x}} \right\|_2^2}}{{\left\| {{{\mathbf{x}}_\textnormal{w}} - {\mathbf{x}}} \right\|_2^2}}.
\end{equation}
The signal-to-noise ratio (SNR) is defined by 
\begin{equation}
\textnormal{SNR}  = 10{\log _{10}}\frac{{\left\| {{{\mathbf{x}}_{\textnormal{w}}}} \right\|_2^2}}{{{\sigma ^2}}},
\end{equation}
where $\sigma^2$ is the variance of additive white Gaussian noise (AWGN). To characterize watermark extraction performance, the detection rate (DR) is defined by
\begin{equation}
\textnormal{DR} = \frac{1}{{2P}}\sum\limits_{i = 0}^{P - 1} {\left| {{\mathbf{w}}(i) - \hat{\mathbf{w}}(i)} \right|}  \times 100\%.
\end{equation}
Parameters are set as follows. $M_0=1024$, $f_1=60$ Hz, $f_2=2800$ Hz, $W=16$, $P=32, 32^2$, $L=W^2$, $\alpha$ is heuristically controlled by ODG values no less than $-1$. During comparison, ODG values are tuned to be similar for fair comparison. In our simulation, four music samples are selected, including male song ($240$ s), female song ($10$ s), violin and piano duet ($10$ s), and electronic music ($10$ s). All the samples audio files have 16-bit quantization and a sampling frequency of $44.1$ kHz.

\subsection{Imperceptibility}
The imperceptibility property of the implemented systems is demonstrated in Fig. \ref{Imp}, where three of the four audio samples with $10$ s duration are used to generate the performance curves, and AWGN with $\textnormal{SNR} = 30$ dB is considered. It can be observed from both sub-figures that the proposed schemes constantly yield better imperceptibility when the DR values are the same. In terms of ODG, the proposed systems could obtain ODG values between $-0.5$ and $0$ with above $90\%$ DRs. Further, the DWR and ODG values of the four audio watermarking systems applied on four audio samples are summarized in Tables \ref{Table1} and \ref{Table2}. It can be seen that performance improvement on imperceptibility is consistent across different samples, and the inability of ISS based methods in controlling imperceptibility is resolved, thanks to localized embedding in selected features. The robustness testing results will be provided in next subsection, also based on the four audio samples, and the corresponding imperceptibility information is as shown in the two tables. We will demonstrate that while the proposed systems could achieve improved imperceptibility, the robustness against several common processing attacks can also be improved.

\begin{table}[!t]
\renewcommand{\arraystretch}{1.2}
\tabcolsep 4mm 	\caption{Imperceptibility of DCT-SS and DCT-ISS methods}
\centering
\vspace{-0.5em}
\begin{tabular}{c|c c| c c}
	\firsthline
   \multirow{ 2}{*}{{Data}}	& \multicolumn{2}{c|}{{DCT-SS}} & \multicolumn{2}{c}{{DCT-ISS}} \\
	\cline{2-3} \cline{4-5}
	 & {DWR (dB)}    & {ODG}  & {DWR (dB)}    & {ODG} \\
	\hline
	Sample 1    & 33.9     & -0.76  & 33.7    & -0.70        \\
	Sample 2    & 34.7     & -0.13  & 32.4    & -0.02     \\
	Sample 3    & 40.4     & -0.40  & 32.8    & -0.90    \\
	Sample 4    & 33.5     & -0.95  & 32.3    & -0.44   \\
	\lasthline
\end{tabular}\label{Table1}
\end{table}

\begin{table}[!t]
\renewcommand{\arraystretch}{1.2}
\tabcolsep 4mm  \caption{Imperceptibility of STFT-SS and STFT-ISS methods}
\centering
\vspace{-0.5em}
\begin{tabular}{c|c c| c c}
	\firsthline
	\multirow{ 2}{*}{{Data}} & \multicolumn{2}{c|}{{STFT-SS}} & \multicolumn{2}{c}{{STFT-ISS}} \\
	\cline{2-3} \cline{4-5}
	   & {DWR (dB)}    & {ODG}  & {DWR (dB)}    & {ODG} \\
	\hline
	Sample 1    & 39.9     & -0.61  & 39.6    & -0.44        \\
	Sample 2    & 40.7     & -0.03  & 42.6    & -0.02     \\
	Sample 3    & 46.4     & -0.43  & 42.1    & -0.40    \\
	Sample 4    & 39.5     & -0.55  & 39.9    & -0.34   \\
	\lasthline
\end{tabular}\label{Table2}
\end{table}

\begin{figure}[!h]
	\centering
	\includegraphics[height=2in,width=3.5in]{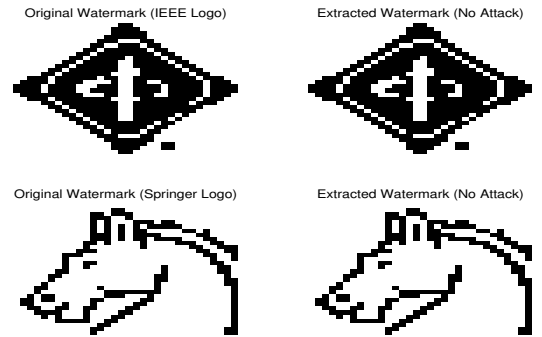}\\[-2em]
	\caption{$32\times 32$-bit watermark logos used with DCT-ISS and STFT-ISS.}
	\label{logos}
\end{figure}

\begin{table*}[!h]
\renewcommand{\arraystretch}{1.45}
\tabcolsep 2.8mm  \caption{DRs (\%) of DCT-SS method under different attacks (first setting)}
\centering
\vspace{-0.5em}
\begin{tabular}{ c |c|c|c|c|c|c }
	\hline
	\multicolumn{2}{c|}{{Attach Type}}  &{Sample 1} & {Sample 2} & {Sample 3} & Sample 4 & {Average} \\ \hline
	\multirow{1}{*}{{Re-Quantization}} 
	& 8 Bit & 84.4   & 96.8  & 78.1  & 93.8 & 88.2     \\
	\hline
	\multirow{2}{*}{{Gaussian Noise}} 
	& 30 dB & 84.4   & 96.8  & 78.1 & 90.6 & 87.5    \\
	& 50 dB & 84.4   & 96.8  & 78.1  & 93.8 & 88.2  \\ \hline
	\multirow{2}{*}{{Amplitude Scal.}} 
	& 1.2  & 84.4   & 96.8  & 78.1 & 93.8& 88.2     \\
	& 1.8 & 84.4   & 96.8  & 78.1  & 93.8 & 88.2       \\
	\hline
	\multirow{2}{*}{{AAC Compression}} 
	& 96 kbps  & 84.4   & 96.8  & 78.1 & 93.8 & 88.2   \\
	& 160 kbps & 84.4  & 96.8  & 78.1  & 93.8 & 88.2  \\
     \hline
	\multirow{2}{*}{{MP3 Compression}} 
	& 64 kbps & 84.4   & 96.8  & 78.1 & 93.8& 88.2   \\
	& 128 kbps & 84.4   & 96.8  & 78.1  & 93.8 & 88.2   \\
    \hline
\end{tabular}\label{Table3}
\end{table*}

\begin{table*}[!h]
	\renewcommand{\arraystretch}{1.45}
\tabcolsep 2.8mm  \caption{DRs (\%) of DCT-ISS method under different attacks (first setting)}
\centering
\vspace{-0.5em}
\begin{tabular}{ c |c|c|c|c|c|c }
\hline
\multicolumn{2}{c|}{{Attach Type}}  &{Sample 1} &{Sample 2} & {Sample 3} & {Sample 4} &{Average} \\ \hline
\multirow{1}{*}{{Re-Quantization}} 
 & 8 Bit & 100  & 100 & 100   & 100 & 100    \\
 \hline
\multirow{2}{*}{Gaussian Noise} 
 & 30 dB & 75.0  & 75.0  & 68.8  & 71.9 & 72.7   \\
 & 50 dB & 100   & 100      & 100& 100 & 100 \\ \hline
\multirow{2}{*}{Amplitude Scal.}
 & 1.2  & 100  & 100 & 100 & 100 & 100    \\
  & 1.8 & 100  & 100   & 100 & 100 & 100      \\
 \hline
\multirow{2}{*}{{AAC Compression}} 
 & 96 kbps  & 96.8  & 96.8   & 96.8  & 96.8 & 96.8   \\
 & 160 kbps & 100   & 100    & 100  & 100 & 100   \\
 \hline
 \multirow{2}{*}{{MP3 Compression}} 
  & 64 kbps & 87.5  & 100  & 93.8 & 100 & 95.3   \\
 & 128 kbps & 100  & 100   & 96.8  & 100 & 99.2   \\
 \hline
\end{tabular}\label{Table4}
\end{table*}

\begin{table*}[!h]
		\renewcommand{\arraystretch}{1.45}
 \tabcolsep 2.8mm \caption{DRs (\%) of STFT-SS method under different attacks (first setting)}
\centering
\vspace{-0.5em}
\begin{tabular}{ c |c|c|c|c|c|c }
\hline
\multicolumn{2}{c|}{{Attach Type}}  & {Sample 1} & {Sample 2} & {Sample 3} & {Sample 4} & {Average} \\ \hline
\multirow{1}{*}{{Re-Quantization}} 
 & 8 Bit & 96.8  & 100  & 81.3  & 96.8 & 93.8    \\
 \hline
\multirow{2}{*}{{Gaussian Noise}} 
 & 30 dB & 93.7  & 100   & 78.1 & 93.8 & 91.4    \\
 & 50 dB & 96.8   & 100   & 81.3 & 96.8 & 93.8 \\ \hline
\multirow{2}{*}{{Amplitude Scal.}} 
 & 1.2  & 96.8   & 100    & 81.3 & 96.8 & 93.8     \\
 & 1.8  & 96.8   & 100    & 81.3 & 96.8 & 93.8       \\
 \hline
\multirow{2}{*}{{AAC Compression}} 
 & 96 kbps  & 96.8   & 100   & 78.1 & 96.8 & 92.9   \\
 & 160 kbps & 96.8   & 100   & 81.3 & 96.8 & 93.8   \\
\hline
\multirow{2}{*}{{MP3 Compression}} 
  & 64 kbps & 96.8   & 100    & 78.1 & 96.8 & 92.9   \\
 & 128 kbps & 96.8   & 100    & 81.3 & 96.8 & 93.8   \\
 \hline
\end{tabular}\label{Table5}
\end{table*}

\begin{table*}[!h]
	\renewcommand{\arraystretch}{1.45}
	\tabcolsep 2.8mm \caption{DRs (\%) of STFT-ISS method under different attacks (first setting)}
	\centering
	\vspace{-0.5em}
	\begin{tabular}{ c |c|c|c|c|c|c }
		\hline
		\multicolumn{2}{c|}{{Attach Type}}  & {Sample 1} & {Sample 2} & {Sample 3} & {Sample 4} & {Average} \\ \hline
		\multirow{1}{*}{{Re-Quantization}} 
 & 8 Bit & 100  & 100  & 100   & 100 & 100    \\
 \hline
\multirow{2}{*}{{Gaussian Noise}} 
 & 30 dB & 96.8   & 93.7  & 96.8  & 87.5 & 93.7    \\
 & 50 dB & 100   & 100      & 100& 100 & 100  \\ \hline
\multirow{2}{*}{{Amplitude Scal.}} 
 & 1.2 & 100  & 100 & 100 & 100 & 100     \\
 & 1.8  & 100  & 100   & 100 & 100 & 100       \\
 \hline
\multirow{2}{*}{{AAC Compression}} 
 & 96 kbps  & 100 & 100  & 100  & 100 & 100  \\
 & 160 kbps & 100   & 100    & 100  & 100 & 100   \\
 \hline
 \multirow{2}{*}{{MP3 Compression}} 
  & 64 kbps & 100  & 100   & 100 & 100 & 100  \\
 & 128 kbps & 100  & 100   & 100  & 100 & 100   \\
 \hline
\end{tabular}\label{Table6}
\end{table*}

\begin{table*}[!h]
\renewcommand{\arraystretch}{1.5}
 \tabcolsep 2.8mm  \caption{DRs (\%) of DCT-ISS and STFT-ISS methods (second setting)}
 \vspace{-0.5em}
 \centering
\begin{tabular}{ cc|c|c|c|c| c}
\hline
\multicolumn{2}{c|}{\multirow{2}{*}{{Attack Type}}}  & \multicolumn{2}{c|}{DCT-ISS} & \multicolumn{2}{c|}{STFT-ISS}  &  \multirow{2}{*}{\shortstack{Average \\ Improvement}} \\
\cline{3-6}
& & $\;\;$IEEE$\;\;$ & Springer & $\;\;$IEEE$\;\;$ & Springer & \\
\hline
\multirow{1}{*}{{Re-Quantization}} 
 & \multicolumn{1}{|c|}{8 Bit} & 94.2  & 95.1 & 100   & 100 & 5.35   \\
 \hline
\multirow{3}{*}{{Gaussian Noise}} 
 & \multicolumn{1}{|c|}{30 dB}& 55.2   & 58.9  & 85.2  & 86.7 & 29.3   \\
 & \multicolumn{1}{|c|}{40 dB} & 70.1   & 69.5   & 94.3 & 95.0  & 24.9    \\
 & \multicolumn{1}{|c|}{50 dB} & 92.6   & 94.5   & 98.6 & 100   & 5.75 \\ \hline
\multirow{2}{*}{{Amplitude Scal.}} 
 & \multicolumn{1}{|c|}{1.2}  & 100   & 100  & 100 & 100   & 0   \\
 & \multicolumn{1}{|c|}{1.8}  & 100  & 100  & 100 & 100   & 0     \\
 \hline
\multirow{3}{*}{{AAC Compression}} 
  & \multicolumn{1}{|c|}{96 kbps} & 75.5   & 73.9  & 85.4  & 82.5 & 9.25   \\
 & \multicolumn{1}{|c|}{128 kbps} & 81.5   & 81.9   & 90.0 & 87.9 & 7.25    \\
 & \multicolumn{1}{|c|}{160 kbps} & 93.4   & 91.9   & 100  & 100  & 7.35 \\
\hline
 \multirow{3}{*}{{MP3 Compression}} 
  & \multicolumn{1}{|c|}{64 kbps} & 64.6  & 75.1   & 85.1  & 86.3  & 15.9  \\
 & \multicolumn{1}{|c|}{128 kbps} & 86.7  & 99.2   & 100  & 100    & 7.05\\
 & \multicolumn{1}{|c|}{192 dbps} & 99.5  & 100    & 100  & 100    & 0.25 \\
 \hline
  & \multicolumn{1}{c}{ODG} & \multicolumn{1}{c}{-0.75} & \multicolumn{1}{c}{-0.76} & \multicolumn{1}{c}{-0.65} & \multicolumn{1}{c}{-0.66} & \\
\end{tabular}\label{Table7}
\end{table*}

\begin{figure*}[!t]
	\centering
	\includegraphics[height=3.7in,width=6.3in]{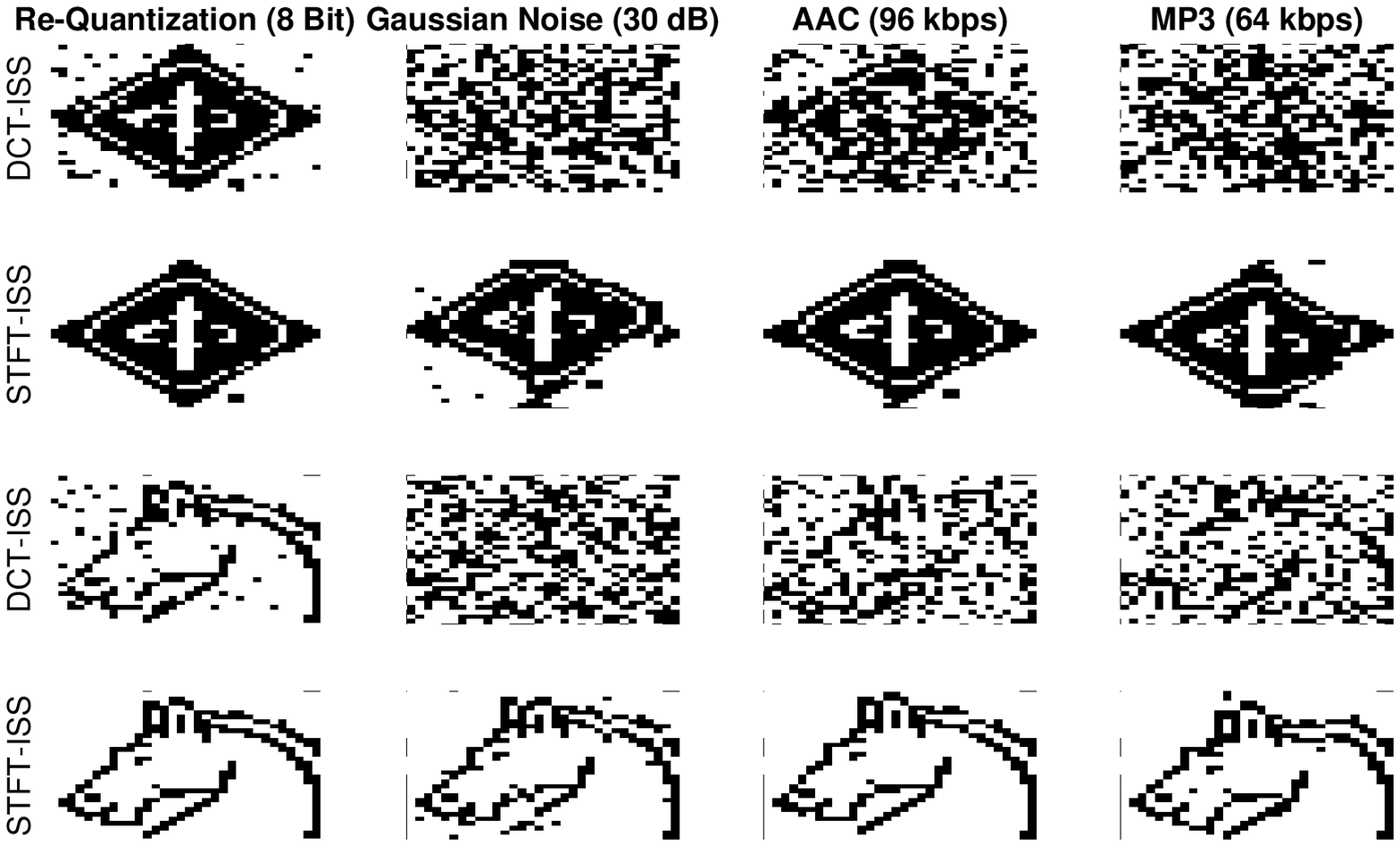}\\[-2em]
	\caption{Recovered $32\times 32$-bit watermark logos (second setting) under different significant attacks used with DCT-ISS and STFT-ISS.}
	\label{logos2}
\end{figure*}

\subsection{Robustness to a Series of Attacks}
The attacks considered in this paper include adding Gaussian noise, amplitude scaling, AAC lossy compression, and MP3 lossy compression, each with several different attack strength settings. Here, we use two sets of watermarks and two sets of sample audio clips. In the first setting, a random binary sequence of $32$ bits is used as the watermark, which is embedded into four audio samples of $10$ seconds. The embedding DWR and ODG values in this setting are given in Tables \ref{Table1} and \ref{Table2}. In the second setting, two graphic logos of $32\times 32$ bits, as shown in Fig. \ref{logos}, are used for watermark embedding, and the $4$-minute audio file is used as the host signal. The ODG values for the second setting are given at the bottom of Table \ref{Table7}. The implemented methods for the second setting are DCT-ISS and STFT-ISS only, for simplicity.

The DRs against several attacks with different attack strength settings for DCT-SS, DCT-ISS, STFT-SS, and STFT-ISS methods are shown in Tables \ref{Table3} to \ref{Table6} respectively. For DCT and STFT based methods respectively, we observe better robustness when ISS is used. This agrees with the classical property of the ISS technique. More importantly, we can see from these tables that the proposed TF feature domain watermarking systems outperform DCT based frequency domain methods for both SS and ISS implementations. The best performance is observed in Table \ref{Table6}, which corresponds to STFT-ISS method. Recall Table \ref{Table2}, in which the DWR and ODG values of STFT-ISS method are very close to, or even slightly better than those obtained from its SS counterpart or DCT based methods. \textbf{Therefore, it could be concluded that the proposed system is able to simultaneously achieve improved imperceptibility and robustness.} 

Finally, we test the two ISS based systems, i.e., DCT-ISS and STFT-ISS in the second setting where IEEE and Springer logos are embedded in a $4$-minute audio clip. The results are shown in Table \ref{Table7}. The improvement of the proposed system against its frequency domain counterpart is consistent across different attacks for both logos.  The recovered IEEE and Springer logos under different significant attacks using the DCT-ISS and the STFT-ISS are shown in Fig. \ref{logos2}. Two observations can be made: the DCT-ISS fails to reconstruct the original logos; Although the STFT-ISS does not achieve completely error-free performance, it can still well recover the shape and content of the logos.

%

\section{Conclusion}
In this paper, we proposed an audio watermarking framework from the perspective of TF analysis. Different from existing schemes, the proposed framework considers the 2D TF representation of host audio signal as the raw signal for watermark embedding. Based on partitioning the 2D TF image into small patches, and selecting patches with lower energy values as features, watermarks are embedded into the vectorized feature patches using SS and ISS mechanisms. Extensive experimental results have been carried out in comparison with the counterpart systems that embed watermark in frequency domain. Consistent performance improvements have been shown via the experimental results, both using random sequences and image logos as watermarks. 

Due to the requirement of feature invariance, the proposed systems are less robust against AWGN attacks. We have noted in Section II.C that using a random indexing key instead of sorting patch energies would be an effective solution to this problem. Future research efforts will be put into this issue. In general, it is worth investigating more efficient TF feature domain audio watermarking methods that could potentially lead to substantially improved robustness against desynchronization attacks. This may possibly be approached via exploring desynchronization invariant features in TF domain and utilizing both SS and QIM based embedding mechanisms.


\end{document}